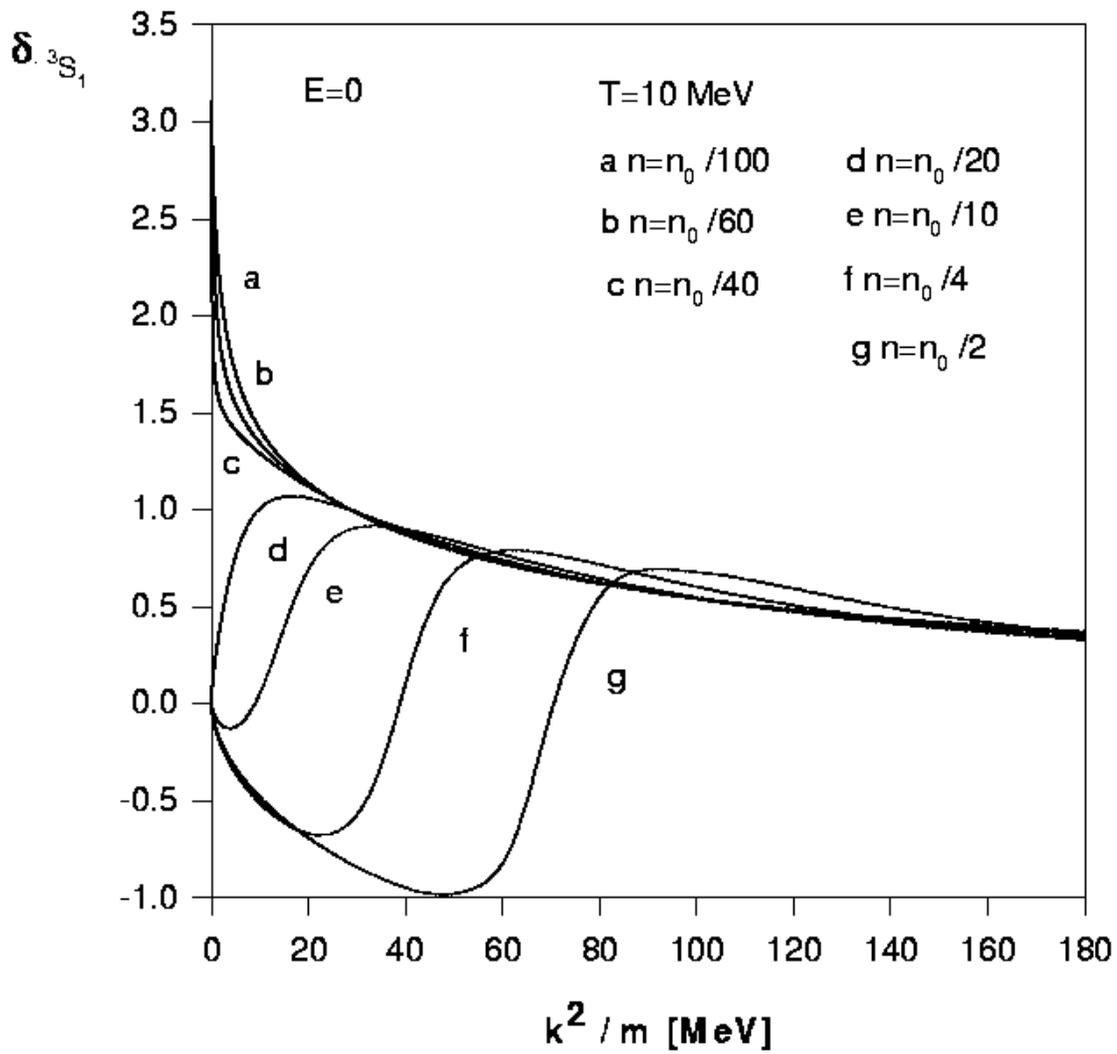

# In-medium two-nucleon properties in high electric fields

K. Morawetz* and G. Röpke

*Max-Planck-Gesellschaft, AG "Theoretische Vielteilchenphysik" an der Universität Rostock,*
*18055 Rostock, Germany*
* *and National Superconducting Cyclotron Laboratory*
*Michigan State University, East Lansing, MI 48824-1321, USA*

(October 25, 1995)

The quantum mechanical two - particle problem is considered in hot dense nuclear matter under the influence of a strong electric field such as the field of the residual nucleus in heavy - ion reactions. A generalized Galitskii-Bethe-Salpeter equation is derived and solved which includes retardation and field effects. Compared with the in-medium properties in the zero-field case, bound states are turned into resonances and the scattering phase shifts are modified. Four effects are observed due to the applied field: (i) A suppression of the Pauli-blocking below nuclear matter densities, (ii) the onset of pairing occurs already at higher temperatures due to the field, (iii) a field dependent finite lifetime of deuterons and (iv) the imaginary part of the quasiparticle self-energy changes its sign for special values of density and temperatures indicating a phase instability. The latter effect may influence the fragmentation processes. The lifetime of deuterons in a strong Coulomb field is given explicitly.

PACS numbers:24.10.Cn,25.70.De,72.20.Ht,25.70.Pq

## I. INTRODUCTION

The problem of Coulomb corrections in heavy - ion reactions raised a considerably interest during the last years. Especially, great attention is paid to the search of signals from the early stage of heavy-ion fragmentation. One of the most intensively studied experimental values is the two - particle correlation function. This observable gives hope to extract information about the process of fragmentation and the situation during the early stage of heavy-ion collisions. Coulomb effects from the residual nucleus are still not well investigated. Here we try to show that this effect may lead to new observable effects. Besides the strong interaction which acts exclusively at short distances of the fragments, the Coulomb field of the residual nucleus is a long range potential and will influence the evolution of the disintegrating nucleus for a longer time interval than the strong interaction.

The most accepted way of describing heavy - ion reactions is to determine transport properties by solving adequate kinetic equations. One of the main ingredients of these equations are the transport cross sections. While a number of works have investigated the density dependence of the cross section and therefore the influence of the medium [1], modification to scattering properties due to the large electric fields have not been considered up to now. In this paper we estimate the influence of the Coulomb field of the residual nucleus the two-particle in-medium properties. The results are of interest for the transport theory of hot expanding nuclear matter in heavy ion collisions.

A pioneering work in the area of field - dependent transport was done in solid state physics [2–7]. One of the problems is the derivation of correct transport equations including the field to arbitrary strengths [3,8–11,7,12]. The application of these high - field phenomena ranges from nonlinear conductivity [13–15] up to quantum coherence phenomena [16]. Among these studies there can be observed two different kinds of effects. The first class consists of kinetic effects resulting from the modification of the single particle distribution function in a strong field. As an example we refer to the stationary nonequilibrium function in strong electric fields [14–19]. A second class of effects are the quantum interference effects [13,20,21]. The latter one is considered in this paper.

A very rough criterion for kinetic field effects is the ratio between energy gain of the charged particle during the mean free path $\lambda$ and the average energy lost during a collision. High field transport is characterized by an increase of the velocity between two successive collisions greater than the mean thermal velocity $\frac{eE}{m}\frac{\lambda}{v_t} > v_t$. The other class of field effects arise by quantum interference. These can be described by a typical time scale $\tau_E = \frac{\hbar}{\varepsilon_E}$, which is determined for one particle properties by $\varepsilon_E = \hbar(\frac{e^2 E^2 \hbar^2}{8m})^{1/3}$.

For two-particle properties [17] we will show that the coherence time is

$$\tau_E = \frac{\hbar}{\left(\frac{E^2\hbar^2}{8}\left(\frac{e_1^2}{m_1} + \frac{e_2^2}{m_2}\right)\right)^{1/3}}. \tag{1}$$

If this time, which can be considered as a collisional broadening time, interferes with the time between two successive collisions one expects new effects of quantum nature [20,21]. The coherence time (1) can be seen from the two-particle phase factor associated with times $t1$ and $t2$. In the c.m for the canonical momentum the phase reads with $T = (t_1 + t_2)/2$ and $\tau = t_1 - t_2$

$$\exp\left(-i\int_{t_2}^{t_1} dt \left(\frac{(\mathbf{p}+e_1\mathbf{E}t)^2}{2m_1} + \frac{(-\mathbf{p}+e_2\mathbf{E}t)^2}{2m_2}\right)\right)$$
$$= \exp\left(-i\left(\left(-\frac{p^2}{2\mu} + \frac{E^2}{2}\left(\frac{e_1^2}{m_1}+\frac{e_2^2}{m_2}\right)T^2\right)\tau - \left(\frac{e_1}{m_1}-\frac{e_2}{m_2}\right)\mathbf{p}\mathbf{E}T\tau + \frac{E^2}{8}\left(\frac{e_1^2}{m_1}+\frac{e_2^2}{m_2}\right)\frac{\tau^3}{3}\right)\right). \tag{2}$$

The term linear in $E$ vanishes for equal $e/m$ and the remaining quadratic term causes an intrinsic field effect. While the first term can be absorbed into redefinition of the energy scale the last term causes an intrinsic field effect with the parameter (1). This is immediately obvious by fouriertransforming to the frequency domain.

Despite the general theoretical interest to describe two-particle properties in high electric fields we focus in this paper on the application to a system likely to be found in heavy ion reactions. To motivate the subsequent investigations we give a simple estimation of expected effects. During the expansion phase of highly excited nuclear matter after a heavy - ion collision the emitted charged particles will be subject to the Coulomb force of the remaining nucleonic system. For a nucleus of charge 100 one has a coherence time of about $\tau_E \approx 100$ fm/c at a distance of 10 fm. This becomes comparable with typical relaxation time scales in the expanding system. Considering the fact that this electric field will not change remarkably during the collision time we describe the effect of the radial Coulomb field in such a way that the spectator particle will get a short push by an effective constant electric field, where the time occurs only as a parameter. Instead of solving the complicated 3 body problem with correct Coulomb wave functions of the residual nucleus we propose to consider the transport during a short time period under the influence of a constant strong electric field. By this way we show how the two - particle properties will be modified owing to the high electric field impulses. This model treatment will be justified by the results discussed in chapter IV, in which we focus on three effects: (i) A suppression of the Pauli-blocking at densities $n_0/40 - n_0/10$, (ii) a pronounced onset of pairing states due to the field in comparison with the zero field case where it takes place at lower temperatures, (iii) the finite lifetime of deuterons due to the electric field and (iv) a modification of the density and temperature region of instability as indicated by the single particle self-energy.

The outline of this paper is as follows. First we derive useful tools for formulating the theory gauge invariant. By the fundamental principle of gauge invariance we will present the in-medium $\mathcal{T}$-matrix equation with the influence of field effects in chapter III. The dynamical behaviour is then discussed briefly. In addition, we give a solution of these field- and medium- dependent Bethe-Salpeter equation for a separable potential of Yamaguchi-type. The interplay between field and medium effects is shown in chapter V, where we focus on the intrinsic field effect of quantum coherence.

## II. GREEN FUNCTION APPROACH

We consider a system of charged nucleons under the influence of an applied constant electric field $\mathbf{E}$. In vector potential gauge $\mathbf{A}(t) = -c\mathbf{E}t$, we have the Hamiltonian

$$H(t) = \sum_{1} \frac{(\mathbf{p_1} - \frac{e_1}{c}\mathbf{A}(t))^2}{2m_1} a_1^+ a_1 + \frac{1}{2}\sum_{1,2,1',2'}\langle 12|V|1'2'\rangle a_{1'}^+ a_{2'}^+ a_2 a_1 \tag{3}$$

with a residual interaction $V$

$$\langle 12|V|1'2'\rangle = \delta(t_1 - t_2)\delta(t_1' - t_2')\delta(t_1 - t_1')V_{ab}(\mathbf{p_1},\mathbf{p_2},\mathbf{p_1'},\mathbf{p_2'})\delta_{\mathbf{p_1}+\mathbf{p_2},\mathbf{p_1'}+\mathbf{p_2'}}$$

which will be assumed separable. In order to describe high field effects and many-particle behavior correctly, it is necessary to use approximations for many particle effects which do not invalidate the resulting equations for arbitrary field strengths. It is known that the gradient expansion with respect to the times corresponds to a linearization in fields [11]. To get an unambiguous way of constructing approximations we have to formulate our theory gauge invariant. This can be done following a procedure known from field theory [22], which has been successfully applied to high field problems in [23]. First we present the idea for the one particle functions [12] and then for two- particle correlations.

In order to describe correlations in highly nonequilibrium situations we define various correlation functions by different products of creation and annihilation operators in the Heisenberg picture

$$G^{>}(1,2) = \frac{1}{i} < \Psi_1(\mathbf{r_1}, t_1)\Psi_2^{+}(\mathbf{r_2}, t_2) >$$

$$G^{<}(1,2) = \mp\frac{1}{i} < \Psi_2^{+}(\mathbf{r_2}, t_2)\Psi_1(\mathbf{r_1}, t_1) > . \tag{4}$$

Here $< ... >$ denotes the average value with the unknown statistical *nonequilibrium* operator $\rho$ and $\mp$ signs the Fermi/Bose system respectively.

## A. Gauge invariant formulation of correlation functions

As mentioned above, special care is required to formulate the theory gauge invariant. This will be achieved in the following way. First we change the space-time coordinates $x_i = (t_i, \mathbf{r}_i)$ into Wigner coordinates $x = x_1 - x_2$ and $X = (x_1 + x_2)/2$. Then we construct gauge invariant correlation functions by a generalized Fourier transform [24]

$$\bar{G}(k, X) = \int d^4x \; e^{\frac{i}{\hbar} \int_{-\frac{1}{2}}^{\frac{1}{2}} d\lambda \; x_\mu [k^\mu + \frac{e}{c} A^\mu(X + \lambda x)]} G(xX). \tag{5}$$

Here $A^\mu = (\Phi, \mathbf{A})$ with $\Phi(\mathbf{R}, T)$ and $\mathbf{A}(\mathbf{R}, T)$ being the scalar and vector potentials, respectively. By gauge transformations of $A^\mu$ and $\Psi$ one establishes immediately that $\bar{G}$ is gauge invariant. While it is possible to apply (5) to the general case, we restrict ourselves here to spatial homogeneous and steady electric fields. In this case (5) reduces to the simple rule [24]

$$\bar{G}(k, X) = \int d^4x \; e^{\frac{i}{\hbar}[x_\mu k^\mu + e \, \mathbf{r}\mathbf{E}T]} G(x, X).$$

Therefore, we must apply the following steps in order to formulate one particle correlation functions gauge-invariant:

1. Fourier transform the difference-variable $\mathbf{r}$ into canonical momentum $\mathbf{p}$

2. Shift the canonical momentum $\mathbf{p}$ to kinematical momentum $\mathbf{k}$ according to $\mathbf{p} = \mathbf{k} - e\,\mathbf{E}\,T$

3. The gauge invariant functions $\bar{G}$ are then given by

$$G(\mathbf{p}, \mathbf{R}, \tau, T) = G(\mathbf{k} - e\mathbf{E}T, \mathbf{R}, \tau, T) = \bar{G}(\mathbf{k}, \mathbf{R}, \tau, T). \tag{6}$$

Next we generalize this method to two particle properties. Using the same Wigner coordinates as above, two-particle correlation functions can be written as

$$G_{ab}(x_1, x_2, X_1, X_2) = \frac{1}{i^2}\langle \Psi_a(X_1 + \frac{x_1}{2})\Psi_a^+(X_1 - \frac{x_1}{2})\Psi_b(X_2 + \frac{x_2}{2})\Psi_b^+(X_2 - \frac{x_2}{2})\rangle \tag{7}$$

Within these coordinates it is possible to introduce the generalized Fourier transform (5) with respect to both difference coordinates

$$\bar{G}_{ab}(k_1, k_2, X_1, X_2) = \int d^4x_1 \int d^4x_2 \; G_{ab}(x_1, x_2, X_1, X_2)$$

$$\times \; \exp\frac{i}{\hbar} \int_{-\frac{1}{2}}^{\frac{1}{2}} d\lambda \; \left( (x_1)_\mu [k_1^\mu + \frac{e_1}{c} A^\mu(X_1 + \lambda x_1)] + (x_2)_\mu [k_2^\mu + \frac{e_2}{c} A^\mu(X_2 + \lambda x_2)] \right). \tag{8}$$

Consequently, the following rule is established to formulate two- particle gauge-invariant properties in the s-channel where $t_1 = t_2, \quad t_1' = t_2', \quad \tau = t_1 - t_1'$ and $\quad T = (t_1 + t_1')/2$:

1. Fourier transform the difference-coordinates $\mathbf{r_1} = \mathbf{x_1} - \mathbf{x_1'}$ and $\mathbf{r_2} = \mathbf{x_2} - \mathbf{x_2'}$ to the canonical momenta $\mathbf{p_1}$ and $\mathbf{p_2}$

2. Shift the canonical momentum to kinematical momentum according to $\mathbf{p_i} = \mathbf{k_i} - e_i \mathbf{E} T$

3. The gauge invariant functions $\bar{G}_{ab}$ are then given by

$$G_{ab}(\mathbf{p_1}, \mathbf{p_2}, \mathbf{R_1}, \mathbf{R_2}, \tau, T) = G_{ab}(\mathbf{k_1} - e_1 \mathbf{E} T, \mathbf{k_2} - e_2 \mathbf{E} T, \mathbf{R_1}, \mathbf{R_2}, \tau, T)$$
$$= \bar{G}_{ab}(\mathbf{k_1}, \mathbf{k_2}, \mathbf{R_1}, \mathbf{R_2}, \tau, T). \quad (9)$$

In the subsequent considerations we will use yet another kind of coordinates which are introduced in the following way [25]

$$< \mathbf{x_1 x_2} | G_{ab}(tt') | \mathbf{x'_1 x'_2} > = \int \frac{d\mathbf{P}}{(2\pi\hbar)^3} \frac{d\mathbf{p}}{(2\pi\hbar)^3} \frac{d\mathbf{p'}}{(2\pi\hbar)^3} e^{\frac{i}{\hbar}\mathbf{P}(\mathbf{x_1}+\mathbf{x_2}-\mathbf{x'_1}-\mathbf{x'_2})/2}$$
$$\times e^{\frac{i}{\hbar}\mathbf{p}(\mathbf{x_1}-\mathbf{x_2}) + i\frac{1}{\hbar}\mathbf{p'}(\mathbf{x'_1}-\mathbf{x'_2})} < \mathbf{p} | G_{ab}(\mathbf{P}, \mathbf{R}, tt') | \mathbf{p'} > \quad (10)$$

where $\mathbf{R} = (\mathbf{x_1} + \mathbf{x_2} + \mathbf{x'_1} + \mathbf{x'_2})/4$. Within these coordinates the gauge invariant form reads

$$< \mathbf{p} | \bar{G}_{ab}(\mathbf{P}, \mathbf{R}, T, \tau) | \mathbf{p'} > = < \mathbf{p} - \frac{e_a - e_b}{2} \mathbf{E} T | G_{ab}(\mathbf{P} - (e_a + e_b)\mathbf{E} T, \mathbf{R}, T, \tau) | \mathbf{p'} + \frac{e_a - e_b}{2} \mathbf{E} T > . \quad (11)$$

These rules are useful in formulating the two-particle properties gauge invariant and will be applied in chapter III.

### B. One-particle properties

Next we give a relation between the different correlation functions of (4) which is not as obvious as in equilibrium case where we have the Kubo-Martin-Schwinger condition holds [26]. Assuming the conventional ansatz, i.e. replace the $\omega$ dependence of the distribution function by their quasi-particle value, the Wigner distribution function $f_W(p, R, T) = \mp i G^<(p, R, \tau = 0, T)$ can be related to the Green's function $G^<$ by

$$G^<(pR\omega T) = \mp i a(pR\omega T) \, f_W(pRT). \quad (12)$$

This is quite good as long as the quasi-particle picture holds and no memory effects play a role. As we noted in chapter II A in the discussion following gauge invariance, the gradient expansion and the formulation of kinetic equations with high fields are basically connected with a careful formulation of retardation times. Therefore, the simple ansatz, called *conventional Ansatz*, will certainly fail.

Another obscure discrepancy is the fact that with the conventional ansatz, one has some minor differences in the resulting collision integrals as compared with results from the density operator technique. With the conventional ansatz, one gets just one-half of all retardation times in the various time arguments [7,24]. This annoying discrepancy remained obscure until the recent work of Lipavsky, *et al.* [27], who established a modified ansatz.

Before we use this ansatz we specify the spectral function including electric fields. For free particles corresponding to parabolic dispersions the gauge invariant spectral function under the influence of constant electric fields is well known [24,4]

$$a_E^0(p,\omega) = 2 \int_0^\infty d\tau \cos\frac{1}{\hbar}\left(\hbar\omega\tau - \frac{p^2}{2m}\tau - \frac{e^2 E^2}{24m}\tau^3\right)$$
$$= \frac{2\pi}{\epsilon_E} \text{Ai}\left(\frac{p^2/2m - \hbar\omega}{\epsilon_E}\right). \quad (13)$$

Here Ai($x$) is the Airy function [28] and we have defined $\epsilon_E = (\hbar^2 e^2 E^2/8m)^{1/3}$. It is instructive to verify that (13) satisfies the frequency sum rule. In the subsequent considerations we will use this spectral function. The time variation of the field $E(t)$ is assumed to be slow compared with $\hbar/\epsilon_E$. Then expression (13) can be used where $E = E(T)$ depends on $T$ as a parameter.

This spectral function can be generalized to interacting systems. Within the quasiparticle picture the quasiparticle energy $\omega = \epsilon_p(R, T)$ is defined by the solution of the dispersion relation

$$\omega\hbar - \frac{p^2}{2m} - Re\Sigma^R(p\omega RT) = 0 \quad (14)$$

where $\Sigma^R$ is the retarded self-energy describing the many-particle influence. Then we find the *field-free* spectral function as a sharp peak near the quasiparticle energies, which are now *independent of* $\omega$. Using Airy transformations [12] it is then possible to construct the field and medium dependent spectral function. In convenient time-representation one finds [4]

$$a_E(k\tau RT) = \exp\left[-\frac{i}{\hbar}\left(\epsilon_k(RT)\tau + \frac{e^2E^2}{24m}\tau^3\right)\right]$$

or in frequency representation

$$a_E(k\omega RT) = \frac{2\pi}{\epsilon_E}\text{Ai}\left[\frac{1}{\epsilon_E}\left(\epsilon_k(RT) - \omega\hbar\right)\right] \quad (15)$$

with $\mathbf{k} = \mathbf{p} + e\mathbf{E}T$ according to (II A). This joint spectral function is a natural generalization of the quasiparticle picture to high-field situations and takes both effects into account: the gauge invariance and the many-particle influence. With the help of the joint spectral function (15) one obtains the ansatz valid for any applied electric field strength [12,17,27] in Hartree-Fock approximation

$$G^<(k\tau RT) = \mp i e^{-\frac{i}{\hbar}\left(\epsilon_k\tau + \frac{e^2E^2}{24m}\tau^3\right)} f_W\left(k - \frac{eE|\tau|}{2}, R, T - \frac{|\tau|}{2}\right) \quad (16)$$

with the quasi particle energies $\epsilon_K$. In order to get more physical insight into this ansatz one can transform to frequency representation [29]

$$G^<(k\omega RT) = \mp i2\int_0^\infty d\tau\cos\frac{1}{\hbar}\left(\hbar\omega\tau - \epsilon_k(R,T)\tau - \frac{e^2E^2}{24m}\tau^3\right) f_W(k - \frac{eE\tau}{2}, R, T - \frac{\tau}{2}). \quad (17)$$

Neglecting the retardation in $f_W$ one recovers the ordinary ansatz (12) with the spectral function (15) [see also (13)]. The generalized ansatz takes into account history by an additional memory. This ansatz is superior to that of Kadanoff-Baym in the case of high external fields in several respects [21]: (i) it has the correct spectral properties, (ii) it is gauge invariant, (iii) it preserves causality, (iv) the quantum kinetic equations derived with Eq.(16) coincide with those obtained from the density matrix technique [7], and (v) it reproduces the Debye-Onsager relaxation effect [20].

## III. $\mathcal{T}$-MATRIX APPROXIMATION

In order to describe short-ranged two-particle interactions it is necessary to introduce the standard approximation of the many-particle theory, the $\mathcal{T}$-matrix approximation [30,25,26]. In the following we derive this approximation with complete time dependence and with the influence of arbitrarily high field strengths. This establishes a generalization of the known Bethe-Salpeter equation. Consequently, we first briefly repeat the general many particle derivation of the ladder approximation. This will be done carefully avoiding gradient expansions in time. In this way we obtain general expressions with complete time convolutions. Then we apply the derived transformation rules to formulate our equations gauge invariant. The Green's functions and the $\mathcal{T}$ matrix are given in gauge invariant forms. This will then be the stage at which approximations are discussed, since we still have reasonable control of field effects.

Considering only binary-collision approximation the causal two-particle Green's functions can be written

$$G_2^{ab}(121'2^+) = G^a(11')G^b(22^+) \mp \delta_{ab}G^a(12^+)G^b(21') + i\int d\bar{1}d\bar{1}'d\bar{2}d\bar{2}'$$
$$[G^a(1\bar{1})G^b(2\bar{2}) \mp \delta_{ab}G^a(1\bar{2})G^b(2\bar{1})] <\bar{1}\bar{2}|\mathcal{T}_{ab}|\bar{1}'\bar{2}'> G^a(\bar{1}'1')G^b(\bar{2}'2^+)$$
$$(18)$$

Here the sum of ladder diagrams is defined as a causal $\mathcal{T}$-Matrix

$$<12|\mathcal{T}_{ab}|1'2'> = V_{ab}(121'2') + i\int d\bar{1}d\bar{2}d3d3'V_{ab}(1233')G^a(3\bar{1})G^b(3'\bar{2})<\bar{1}\bar{2}|\mathcal{T}_{ab}|1'2'>. \quad (19)$$

Since the interacting potential is assumed to be local in time, we can simplify the general equations. From the definition of the $\mathcal{T}$-matrix (19) we get

$$< 12|\mathcal{T}|1'2' > = < x_1 x_2 t_1|\mathcal{T}|x'_1 x'_2 t'_1 > \delta(t_1 - t_2)\delta(t'_1 - t'_2).$$

With the abbreviation

$$< x_1 x_2 t|\mathcal{G}^{ab}|\bar{x}_1 \bar{x}_2 \bar{t} > = G^a(x_1 t \bar{x}_1 \bar{t}) G^b(x_2 t \bar{x}_2 \bar{t}) \qquad (20)$$

we get the causal $\mathcal{T}$-matrix

$$< x_1 x_2 t|\mathcal{T}_{ab}|x'_1 x'_2 t' > = V_{ab}(x_1, x_2, x'_1, x'_2)\delta(t - t')\, 2\, \delta(x_1 + x_2 - x'_1 - x'_2)$$
$$+ i \int d\bar{x}_1 d\bar{x}_2 dx_3 dx'_3 V_{ab}(x_1, x_2, x_3, x'_3) < x_3 x'_3 t|\mathcal{G}_{ab}|\bar{x}_1 \bar{x}_2 \bar{t} >< \bar{x}_1 \bar{x}_2 \bar{t}|\mathcal{T}_{ab}|x'_1 x'_2 t' > .$$
$$(21)$$

Now we can apply the Lengreth-Wilkins rules [31,29] and find the relations in operator notation

$$\mathcal{T}^{\stackrel{>}{<}} = iV\mathcal{G}^R \mathcal{T}^{\stackrel{>}{<}} + iV\mathcal{G}^{\stackrel{>}{<}} \mathcal{T}^A$$
$$\mathcal{T}^{R/A} = V + iV\mathcal{G}^{R/A} \mathcal{T}^{R/A}. \qquad (22)$$

By replacing the operator $1 - iV\mathcal{G}^R$ in the first equation with the help of the second one we derive the generalized optical theorem [30,32,33]

$$< x_1 x_2 t|\mathcal{T}_{ab}^{\stackrel{>}{<}}|x'_1 x'_2 t' > \;=\; \int d\bar{x}_1 d\bar{x}'_1 d\bar{x}_2 d\bar{x}'_2 d\bar{t} d\bar{t}' < x_1 x_2 t|\mathcal{T}_{ab}^R|\bar{x}_1 \bar{x}_2 \bar{t} >$$
$$\times \; < \bar{x}_1 \bar{x}_2 \bar{t}|\mathcal{G}_{ab}^{\stackrel{>}{<}}|\bar{x}'_1 \bar{x}'_2 \bar{t}' >< \bar{x}'_1 \bar{x}'_2 \bar{t}'|\mathcal{T}_{ab}^A|x'_1 x'_2 t' > . \qquad (23)$$

From the $\mathcal{T}$-matrix equations (22) we get for the retarded one within the coordinates (10)

$$< p|\mathcal{T}_{ab}^R(PRtt')|p' > = V_{ab}(p, p')\delta(t - t') + i \int \frac{d^3\bar{p}}{(2\pi\hbar)^3} \int_{-\infty}^{t} d\bar{t} V_{ab}(p, \bar{p}) < \bar{p}|\mathcal{T}_{ab}^R(PR\bar{t}t')|p' >$$
$$\times \left[ G_a^>(\frac{P}{2} - \bar{p}, R, t\bar{t}) G_b^>(\frac{P}{2} + \bar{p}, R, t\bar{t}) - G_a^<(\frac{P}{2} - \bar{p}, R, t\bar{t}) G_b^<(\frac{P}{2} + \bar{p}, R, t\bar{t}) \right]. \qquad (24)$$

This system of equations was derived assuming small spatial micro-fluctuations in comparison to macro-observables $R$. Thus, gradient expansions in space can be applied. It is important to remark that the equations up to now contain the exact time behavior for both microscopic times as well as center-of-mass times. Now we formulate these equations gauge invariant.

We change the time variables according to $\bar{t} = t + \bar{\tau}, T = (t + t')/2, \tau = t - t'$ and apply the rules (11) to formulate the quantum mechanical two-particle problem gauge invariant. Then Eq. (24) becomes

$$< p + \frac{e_a - e_b}{2} ET|\bar{\mathcal{T}}_{ab}^R(K_a + K_b, R, T, \tau)|p' - \frac{e_a - e_b}{2} ET > = V_{ab}(p, p')\delta(\tau)$$
$$+ i \int \frac{d^3\bar{p}}{(2\pi\hbar)^3} \int_{-\tau}^{0} d\bar{\tau} V_{ab}(p, \bar{p}) \left[ \bar{G}_a^>(K_a - \bar{p} + e_a E \frac{\bar{\tau} + \tau}{2}, R, -\bar{\tau}, T + \frac{\bar{\tau} + \tau}{2}) \right.$$
$$\times \bar{G}_b^>(K_b + \bar{p} + e_b E \frac{\bar{\tau} + \tau}{2}, R, -\bar{\tau}, T + \frac{\bar{\tau} + \tau}{2}) - \bar{G}_a^< \bar{G}_b^< \Big]$$
$$\times \langle \bar{p} + \frac{e_a - e_b}{2} E(T + \frac{\bar{\tau}}{2})|\bar{\mathcal{T}}_{ab}^R(K_a + K_b + (e_a + e_b)E\frac{\bar{\tau}}{2}, R, T + \frac{\bar{\tau}}{2}, \tau + \bar{\tau})|p' - \frac{e_a - e_b}{2} E(T + \frac{\bar{\tau}}{2})\rangle.$$
$$(25)$$

The abbreviations $K_a = \frac{P}{2} + e_a ET$ and $K_b = \frac{P}{2} + e_b ET$ have been introduced for compactness.

Now we assume in agreement with our model of spatial and temporal homogeneous fields, that the macroscopic time $T$ is much larger than the microscopic difference time $\tau$. This is justified if we consider the macroscopic time scales is in orders of the two-particle coherence time $\tau_E = 100 fm/c$ [see introduction and also the discussion after (29)]. Because $\bar{\tau}$ is smaller than $\tau$ due to the integration limits, we see that $T >> \bar{\tau}$ and $T >> \tau$. In energy domain this means that we demand

$$|\hbar \frac{\partial}{\partial \omega} \frac{\partial}{\partial T}| << 1.$$

For typical energies about $25 MeV$ this means that we restrict our considerations to macroscopic time $T >> 4 fm/c$. From this we expand $\bar{\tau}$ and $\tau$ where they occur together with the macroscopic time $T$ in Eq. (25) and obtain after introducing the ansatz (16) to zero order

$$< p + \frac{e_a - e_b}{2} ET |\bar{\mathcal{T}}_{ab}^R (K_a + K_b, T, \tau)| p' - \frac{e_a - e_b}{2} ET > = V_{ab}(p, p') \delta(\tau)$$
$$+ i \int \frac{d^3 \bar{p}}{(2\pi\hbar)^3} V_{ab}(\bar{p}, p) \left(1 - f_W^a (K_a - \bar{p}, T) - f_W^b (K_b + \bar{p}, T)\right)$$
$$\times \int_0^\tau dx < \bar{p} + \frac{e_a - e_b}{2} ET |\bar{\mathcal{T}}_{ab}^R (K_a + K_b, x, T)| p' - \frac{e_a - e_b}{2} ET >$$
$$\times e^{i \left( \left[\frac{(K_a - \bar{p})^2}{2m_a} + \frac{(K_b + \bar{p})^2}{2m_b}\right](x-\tau) + \frac{E^2}{24}(\frac{e_a^2}{m_a} + \frac{e_b^2}{m_b})(x-\tau)^3 \right)}$$
(26)

Here we changed $\bar{\tau} = x - \tau$. The energy arguments of the distribution functions are the quasi-particle energies. In appendix A and [17,12] a slightly more general result is found including retardation effects. Before we solve this equation in the next chapter we give the link to known results. In the field-free limit one recovers the standard result of the Bethe-Salpeter equation. To see this it is possible to write (26) in a more familiar way

$$< p + \frac{e_a - e_b}{2} ET |\bar{\mathcal{T}}_{ab}^R (K_a + K_b, \omega, T)| p' - \frac{e_a - e_b}{2} ET > = V_{ab}(p, p') + \int \frac{d\bar{\mathbf{p}}}{(2\pi\hbar)^3} V_{ab}(\bar{p}, p)$$
$$\times < \bar{p} + \frac{e_a - e_b}{2} ET |\bar{\mathcal{T}}^R (K_a + K_b, \omega, T)| p' - \frac{e_a - e_b}{2} ET > \Phi(\frac{(K_a - \bar{p})^2}{2m_a} + \frac{(K_b + \bar{p})^2}{2m_b} - \bar{\omega})$$
$$\times \left(1 - f_W^a (K_a - \bar{p}, T) - f_W^b (K_b + \bar{p}, T)\right).$$
(27)

The function $\Phi$ is given in terms of Airy-functions [28]

$$\Phi(z) = \frac{\pi}{\lambda_E} \left( \text{Gi}(\frac{z}{\lambda_E}) + i \text{Ai}(\frac{z}{\lambda_E}) \right).$$
(28)

Here the constant is $\lambda_E = \hbar \nu_E$ with the frequency $\nu_E$ introduced as

$$\nu_E = \frac{1}{\hbar} \left( \frac{E^2 \hbar^2}{8} (\frac{e_a^2}{m_a} + \frac{e_b^2}{m_b}) \right)^{\frac{1}{3}}.$$
(29)

As can be seen, the function $\Phi$ reduces to $1/(z + i\eta)$ in the field free case, where $\nu_E$ tends to zero, such that the known Bethe-Salpeter equation occurs as the standard $\mathcal{T}$-matrix equation [1,25,30,26,32]. Therefore we can interpret the inverse frequency $\tau_E = \nu_E^{-1}$ as the characteristic timescale describing the intra-collisional field effect. Similar time scales have been analyzed in detail in [34,35] for an electron-phonon system. The effect of finite collision duration is also presented in [36,6], where similar expressions for are discussed with respect to the band structure effects in a semiconductor.

Without fields, the Bethe-Salpeter equation (27) with the factor $1 \mp f \mp f = (1 \mp f)(1 \mp f) - ff$ includes, in principle, more than the Bloch-de Dominicis equation [37] with a respective factor $(1 \mp f)(1 \mp f)$, by allowing for both intermediate particle – particle and hole – hole, excitations. In the zero-temperature limit for Fermions Eq. (27) corresponds to the Galitskii [38] equation, while the Bloch-de Dominicis equation corresponds to the Bethe – Goldstone [39] equation. Differences between the equations have been discussed in detail in the literature ( for references, see [30] ).

## IV. SOLUTION OF THE $\mathcal{T}$-MATRIX EQUATION

In consequence of the assumed separable potential of rank one

$$V_{ab}(p, p') = \lambda g(p) g(p')$$

we can represent the retarded $\mathcal{T}$-matrix as

$$\langle \mathbf{p}|\bar{\mathcal{T}}_{ab}^R(\mathbf{P},\mathbf{R},T,\tau)|\mathbf{p}'\rangle = \lambda g(\mathbf{p}' - \frac{e_a - e_b}{2}\mathbf{E}T)g(\mathbf{p} + \frac{e_a - e_b}{2}\mathbf{E}T)\mathcal{S}(\mathbf{P} - (e_1 + e_2)\mathbf{E}T,\mathbf{R},\tau,T). \tag{30}$$

Introducing this into eq (27) we can solve the equation for $\mathcal{S}$ by Fourier-transformation in difference time $\tau$. The solution of the $\mathcal{T}$-matrix equation (26) reads

$$\langle \mathbf{p}|\bar{\mathcal{T}}_{ab}(\mathbf{P},\mathbf{R},\omega,T)^R|\mathbf{p}'\rangle = \frac{\lambda g(\mathbf{p}' - \frac{e_a - e_b}{2}\mathbf{E}T)g(\mathbf{p} + \frac{e_a - e_b}{2}\mathbf{E}T)}{1 - \lambda J(\mathbf{P},\mathbf{R},\mathbf{E},\omega,T)} \tag{31}$$

with

$$J(\mathbf{P},\mathbf{R},\mathbf{E},\omega,\mathbf{T})$$
$$= \int \frac{d\bar{\mathbf{p}}}{(2\pi\hbar)^3} g(\bar{\mathbf{p}} + \frac{e_a - e_b}{2}\mathbf{E}T)^2 \left(1 - f_W^a(\frac{\mathbf{P}}{2} - \bar{\mathbf{p}},\mathbf{R},T) - f_W^b(\frac{\mathbf{P}}{2} + \bar{\mathbf{p}},\mathbf{R},T)\right)\Phi(\frac{P^2}{4m} + \frac{\bar{p}^2}{m} - \omega). \tag{32}$$

The function $\Phi$ is given by (28). In the previous equations the mass difference between the particle has been neglected. The more general expression is obvious.

Equation (31) represents the main result of this paper. It gives the solution of the medium dependent Bethe-Salpeter or $\mathcal{T}$-matrix equation in high fields including the medium effects via Pauli blocking factors. Similar investigations were performed in [7] within the framework of a random-layer model for semiconductors where the equations are less involved and the scattering $\mathcal{T}$-matrix was used.

Before we give numerical results in the next chapter, let us qualitatively discuss the physical content of (31). The field effects arise in two different ways. First, for different charged particles $e_a \neq e_b$, the applied field will decrease the $\mathcal{T}$-matrix due to the coupling terms $(e_a - e_b)\mathbf{E}T$ and therefore will destroy any two particle correlation with increasing time. This is basically the fact that the two particles will be accelerated in different ways and lose coherence. In what follows we estimate the different time scales and show that the macroscopic time $T$ is separated from the microscopic time $\tau$ in the nuclear matter situation considered. The two particles become uncorrelated after a time $T$ when the mean value of the momentum $p < |e_a - e_b|E_cT$ corresponding to (31). Equating the latter expression we define the correlation time $T_{\text{corr}} = p/(|e_a - e_b|E_c)$. In a nuclear matter situation the effective radial Coulomb field of a nucleus with charge $Z_n$ acting on a spectator is given by

$$E_c = \frac{e}{4\pi\epsilon_o}\frac{Z_n}{\bar{r}^2} \approx \frac{Z_n}{\bar{r}^2} \times 10^6 \times \text{V} \times \text{fm}.$$

This means an elementary charged particle will feel a field of 100 MV/fm in a distance of 1 fm of the residual, charge-100 nucleous. The correlation time $T_{\text{corr}}$ is therefore $T_{\text{corr}} = 2\frac{fm}{c}(\frac{r}{fm})^2$ for nuclear mater conditions. Now we return to the approximation which leads from eq (25) to (26). The macroscopic time $T$ of the two-particle correlations are determined by the orders of $T_{corr}$. The microscopic time $\tau$ of eq. (25) is of the order of the time the particle needs to move the difference coordinate r. For this distance r the particle needs $T = 3\frac{fm}{c}\frac{r}{fm}$. Therefore the macroscopic time increases quadratically with the distance whereas the microscopic time is proportional to the distance from the residual nucleous. This means that our assumption of $T >> \tau$ in deriving (26) from (25) is well justified. A generalization including retardation effects is given in the appendix A.

For further numerical illustration of the content of equation (31) we will limit the discussion to equal charged particles like p-p correlations or assume for the triplet channel a sudden enough push of the electric field in order to establish stationary conditions – where only the intrinsic field effect is important. This second effect, which we focus on, is to be observed mainly for equal charged particles where the time dependence of the $T$ matrix vanishes. The remaining coherence effect is condensed in the function $\Phi$ in (28). In the field free case the function $\Phi$ reduces to $1/(z + i\epsilon)$ such that the known Bethe-Salpeter equation occurs [1]. This effect is a pure quantum mechanical one and is not possible to understand by classically. In the classical case one would expect that the correlation between two equally charged particles will not be changed by a constant electric field. Here we see that the correlations between two quantum objects can be altered by an applied electric field. This reflects the nonlocality of the scattering process. In semiconductor physics this effect is sometimes called intra-collisional field effect [24,34].

## V. NUMERICAL RESULTS

For the purpose of exploratory calculation of the expected effects we choose a model for the interaction of Yamaguchi-type [40]

$$V_{ab}(p, p') = \lambda g(p) g(p')$$

where

$$g(p) = \frac{1}{p^2 + \beta^2}.$$

The parameters $\lambda$ and $\beta$ are chosen to reproduce the bound state energy of the deuteron and the scattering length for the triplet phase.

In eq. (32) we performed the azimuthal integration and the remaining two dimensional integral was done numerically. It involves fast oscillating functions and is in general dependent on 3 angles, i.e. $(p, E), (p', E)$ and $(K, E)$. In our case we consider the quasistationary condition, e.g. for equal charged particles, wherein the first two angular dependence vanish and only $(K, E)$ remains. In the following we restrict to longitudinal fields with respect to the center of mass momentum $K$, which represents the maximal expected field effect. Due to the complicated field coupling a partial wave decomposition would be senseless here. Therefore we did not use angular averaged Pauli-blocking factors, but considered them completely.

### A. Scattering properties

First we give the numerical solution of the $\mathcal{T}$-matrix at positive energies. There the $\mathcal{T}$-matrix becomes a complex valued function because $J$ has a nonvanishing imaginary part. Similar to the low density limit, the on-shell scattering is characterized by phase shifts. Starting from the solution of the $\mathcal{T}$-matrix in our model we consider the ratio

$$\tan \delta(p, n, T, E) = \frac{\mathrm{Im} < \mathcal{T} >}{\mathrm{Re} < \mathcal{T} >}, \tag{33}$$

which carries certain information about the $\mathcal{T}$-matrix and which has a simple interpretation as the momentum, density and temperature-dependent phase shift in the field free case [1]. In the considered field dependent case, this is a formal parameterization of the $\mathcal{T}$ matrix. The interpretation as phase shifts known from ordinary scattering theory are not obvious anymore. This is connected with the fact that one does not have plane waves as incoming particles. In high field case the incoming plane wave is modified to an Airy-function. The quantum mechanical scattering is therefore much more complicated. But the above definition can be understood as a quantitative measure of the medium and field dependent $\mathcal{T}$-matrix and can be compared to the known phase shifts from field free solutions.

For instance in figure 1 we plot the density dependent phase shifts for a temperature of 10 MeV versus the relative energy of two nucleons having a total momentum $k = 0$ for the field- free case. All phase shifts were obtained with the numerical representation of the $\delta$-function and the principal value, to which the expression (28) should collapse for vanishing fields. This highly oscillating function for small fields serves as a check for the numerics. The result agrees with the former ones in [41,1] as it should. For energies $k^2/m < \mu$ the Pauli blocking factor $(1 - f_W^a - f_W^b)$ becomes negative, so that this term in the Bethe-Salpeter equation acts like a repulsive force which gives rise to negative phase shifts at special densities [1]. This is to be seen for densities above $n_0/10$.

From the behaviour of the phase shifts at $k^2 = 0$ one concludes that in the low density limit the triplet state possesses a bound state following Levinson's theorem. As can be deduced from the figure this bound state vanishes between the densities $n_0/20$ and $n_0/30$ turning the phase shift from $\pi$ to $0$. This behaviour will become more transparent when the bound states are considered explicitly in the next paragraph.

For special poles $\mathrm{Re}(1 - J(\omega = 2\mu)) = 0$ the $\mathcal{T}$-matrix approximation of the equation of state breaks down and the onset of BCS state can be observed [1,42]. In Fig. 4 the phase shifts are plotted for different temperatures at a fixed density. One sees clearly the enhancement of the resonance at $k^2/m = 2\mu$ indicating the onset of superfluity and the breakdown of the $\mathcal{T}-$ matrix approach. Further on we concentrate on temperatures above this critical value.

In figure 2 we plot the same temperature and density dependent phase shifts but let an electric field act on the two particles. Here we introduce a scaling field-strength of

$$E_c = \frac{\hbar^2}{m} \sqrt{\frac{8}{Z_a^2 + Z_b^2}} \times \mathrm{e}^{-1} \times \mathrm{fm}^{-3}, \tag{34}$$

where $Z_a, Z_b$ are the charges of the correlated particles. As can be seen we no longer have bound states as Levinson's theorem would suggest. This becomes more obvious in the figure 3 a where the low density phase shifts are plotted with and without fields.

The second observation is that the Pauli-blocking effect becomes suppressed due to the field. Whereas in the field-free case the effective potential becomes repulsive at higher densities we see in figure 3 b that for densities around $n_0/50 - n_0/20$, where the field free shift already changes the sign, Pauli-blocking is less remarkable within electric fields. On account of the field additional phase space becomes available for scattering which is not accessible without fields essentially due to Pauli-blocking. This can be understood by the higly oscillating spectral function which leads to a collisional broadening resulting into a finite probability to scatter into higher energy states than the exact energy conservation would allow. This effect is not vanishing by averaging because of the highly asymmetric shape of the spectral function. Therefore it remains an additional probability to scatter beyond the Fermi energy due to the field.

A third observation is that the onset of the singularity in the $\mathcal{T}$-matrix due to the pairing is pronounced in a similar way as it is in the field-free situation, but occurs already at higher temperatures. As discussed in [1] the pole in the $\mathcal{T}$-matrix, which is located at $k^2/m = 2\mu$ in the field free and low density limit, gives rise to the onset of BCS states. The onset of this singularity is to be seen in the resonance around $k^2/m \equiv 2\mu$ which becomes more pronounced at low temperatures. In Fig. 4 we plot the phase shifts above and below the critical temperature for different densities. The sharp increase near $k^2/m \equiv 2\mu$ is demonstrated. The critical temperature is around 5 MeV. The same abrupt onset of transition to superfluity is now observed with an electric field, but already at higher temperatures. Fig. 3 c shows the phase shifts for a density $n_0/10$ and a temperature of $T = 10$ MeV where the field free shift changes sign and starts to show a resonance. In the field dependent case the resonance is very pronounced like it is for lower temperatures in the field free case. Therefore we believe that the pairing area in the temperature - density plane is enlarged due to field effects. This effect is very puzzling in the sense that from ordinary transport theory one expects an increase of the particle temperature and a destruction of coherence due to the field. Here we found quite the opposite, namely, for the two- particle problem the coherence phenomena leads to a higher pairing temperature, which would corresponds to lower kinetic temperature without coherence. This means the particles become ordered already at higher temperatures due to the field. This is understandable in the way that the coherence is built up. In other words, we obtain an enforced onset of coherence in a constant external field by the quantum mechanical nature of the scattering process.

### B. Bound states in high field

Turning to the solution of the $\mathcal{T}$-matrix with negative energies we can identify the poles as bound states in the system. In figure 5a we give the imaginary versus the real parts of the pole in dependence of the density at 10MeV. One sees that the bound state vanishes around $n_0/20 - n_0/30$. At these densities the imaginary part is zero translating into long living correlations, i.e. bound states.

In the next figure 5b we give a plot of the imaginary versus the real parts of the poles at applied field strengths of $20 E_c$ corresponding to the scattering data in the last paragraph for the same varying values of density. We see that the poles acquire imaginary parts indicating the finite lifetime of the correlation. It is interesting to notice the nonlinear behaviour of the correlation lifetime with increasing density. First we see that with higher density the life-time increases slightly. At density regions where the Pauli-blocking effect will cause a change to a repulsive interaction, we see that the lifetime decreases with higher densities in the field-free case. The same nonlinear behaviour is observed in the next plot of real and imaginary parts vs. the applied electric field in Fig 6 for a low density limit. The energy of the correlation is reduced from the deuteron binding energy $-2.22$ MeV with increasing fields. At the same time the imaginary part rises sharply indicating that the bound state becomes a resonance. At a field strength of $15 E_c$ the energy of the resonance turns back and crosses the real axes at a critical field strength of $20 E_c$. Correspondingly the imaginary part has a maximum indicating the shortest lifetime. The higher field behaviour is characterized by oscillations up to unstable behaviour, which is given by negative damping.

The oscillating behaviour becomes more observable if one considers a density where the bound state is absent in the field free case. For a chosen density of $n_0/4$ in Fig 7 corresponding to the density in case $d$ of Fig 1, we see that the pole moves in the complex plane crossing the real axes at higher critical fields than it is the case at lower densities. It is remarkable to observe that the damping in Fig. 7 is negative for small fields. This means that the system is unstable in situations where no bound states occur in field-free situations. The resonances are such that the correlations give rise to unstable behaviour in small fields, which may influence the fragmentation process.

## VI. SUMMARY

In the present paper we give have studied the problem of Coulomb signals from the early stage of heavy ion collisions. We have found that the influence of the Coulomb field of the residual nucleus on nucleonic correlations is important and can be estimated by a model of correlations which gets a Coulomb " push " in the early stage of evaporation or fragmentation. This modelled push of a constant field allows the formulation of the original three particle problem as a two-particle problem with an external field. By an explicit gauge invariant formulation of the nonequilibrium $\mathcal{T}$-matrix we derive the Bethe-Salpeter equation with exact time evolution generalized to the influence on external electric field of arbitrary strength.

Using a simple Yamaguchi-type of interaction we presented the solution of the Bethe-Salpeter equation with medium and field dependence. Field-free results reduce to results of earlier publications. An interesting interplay was found between medium and field effects. We observed that the two-particle bound states become resonances with corresponding field dependent lifetimes. A minimal lifetime occurs in dependence on field strengths. These bound states with finite lifetimes are shifted towards positive values of energy which can be considered as the transition to scattering states. This indicates a scattering state as a two-particle correlation in the medium and field.

Suppression of the phase shifts due to Pauli blocking is lowered by the applied field and can be understood in that the available phase space for scattering is enlarged by an applied external field. Furthermore, the onset of superfluidity as a resonance in the scattering phase shift is seen to become significant already at higher temperatures in comparison to the field-free case. This is interpreted as the enforced onset of coherence due to the external field.

By exploratory numerical solutions of the complex bound state energies we observe unstable behaviour at smaller field strengths and high densities which corresponds to the field-free situation where the bound states are already absent due to Mott transitions.

## ACKNOWLEDGMENTS

The authors are indebted to Kevin Haglin for reading the manuscript. This work was supported by a grant (K.M) of the German Academic Exchange Service.

## APPENDIX A: BETHE-SALPETER EQUATION WITH RETARDATION EFFECTS

Here we like to present a less restrictive approximation to the exact equation (25), which was carried out in deriving eq. (26). It will lead to the inclusion of retardation effects.

Instead of neglecting all difference times in (25) as compared to the macroscopic time $T$ we now concentrate to the exact phase factor in (25). We introduce the ansatz (16) and the two particle propagator in (25) takes the following form

$$\begin{aligned} G_a^> G_b^> - G_a^< G_b^< = \\ \exp\left(i\left[\frac{(\mathbf{K_a} - \bar{\mathbf{p}} + e_a \mathbf{E}(\frac{\tau + \bar{\tau}}{2}))^2}{2m_a} + \frac{(\mathbf{K_b} + \bar{\mathbf{p}} + e_b \mathbf{E}(\frac{\tau + \bar{\tau}}{2}))^2}{2m_b}\right]\bar{\tau} + i\frac{E^2}{24}(\frac{e_a^2}{m_a} + \frac{e_b^2}{m_b})\bar{\tau}^3\right) \\ \times \left(1 - f_W^a(K_a - \bar{p} - e_a E(\frac{\tau}{2} + \bar{\tau}), T + \frac{\tau}{2} + \bar{\tau}) - f_W^b(K_b + \bar{p} - e_b E(\frac{\tau}{2} + \bar{\tau}), T + \frac{\tau}{2} + \bar{\tau})\right). \end{aligned} \quad (A1)$$

If one now introduces (A1) into (25) one sees an complicated dependence on the difference time $\tau$ as well as the integration time $\bar{\tau}$.

Besides the $\bar{\tau}$ dependence, which yields an oscillating function of Airy-type, there is an additional dependence on $T + \frac{\tau + \bar{\tau}}{2}$ in the arguments of the quasiparticle energies. This would yield additional oscillations besides the one already described. Therefore we restrict the attention to the mean oscillating function which was treated in chapter III. This means we should ensure that $T + \frac{\tau + \bar{\tau}}{2} \to T$ in the phase factor. This can be done in a saddle point like manner by replacing $\bar{\tau} \to -\tau$ in the remaining functions under the time integral. We obtain instead of (26) the slightly more general result

$$< p + \frac{e_a - e_b}{2}ET|\bar{\mathcal{T}}_{ab}^R(K_a + K_b, \tau T)|p' - \frac{e_a - e_b}{2}ET> = V_{ab}(p, p')\delta(\tau) + i\int \frac{d^3\bar{p}}{(2\pi\hbar)^3}V_{ab}(\bar{p}, p)$$

$$\times \int_0^\tau dx < \bar{p} + \frac{e_a - e_b}{2}ET|\bar{\mathcal{T}}_{ab}^R(K_a + K_b + \frac{e_a + e_b}{2}E(\tau), x, T - \frac{\tau}{2})|p' - \frac{e_a - e_b}{2}ET>$$

$$\times \quad e^{i\left(\left[\frac{(K_a-\bar{p})^2}{2m_a}+\frac{(K_b+\bar{p})^2}{2m_b}\right](x-\tau)+\frac{E^2}{24}(\frac{e_a^2}{m_a}+\frac{e_b^2}{m_b})(x-\tau)^3\right)}$$
$$\times \quad \left(1-f_W^a(K_a-\bar{p}-e_aE\frac{\tau}{2},T-\frac{\tau}{2})-f_W^b(K_b+\bar{p}-e_bE\frac{\tau}{2},T-\frac{\tau}{2})\right). \tag{A2}$$

In this result we account for retardation which leads to different dependence of the difference time $\tau$. The energy arguments of the distribution functions are the quasiparticle energies including off-shell effects by memory [29]. In the same manner as was done in deriving (27) we can express (A2) in a more familiar way using the Airy-transformation [43,23]

$$<p^*|\bar{T}_{ab}^R(K_a+K_b,\tau,T)|p'^*> = V_{ab}(p,p') + i\int\frac{d\bar{\mathbf{p}}}{(2\pi\hbar)^3}V_{ab}(\bar{p},p)$$
$$\times \quad \int\frac{d\Omega}{2\pi}\frac{\mathcal{AI}\left(e^{i\omega\tau}<\bar{p}^*|\bar{T}^R(K_a+K_b-\frac{e_a+e_b}{2}E\tau,\omega,T-\frac{\tau}{2}|p'^*>;\omega\right)(\Omega)}{\frac{(K_a-\bar{p})^2}{2m_a}+\frac{(K_b+\bar{p})^2}{2m_b}-\Omega-i\epsilon}$$
$$\times \quad \left(1-f_W^a(K_a-\bar{p}-e_aE\frac{\tau}{2},T-\frac{\tau}{2})-f_W^b(K_b+\bar{p}-e_bE\frac{\tau}{2},T-\frac{\tau}{2})\right) \tag{A3}$$

with $p^*=p+\frac{e_a-e_b}{2}ET$, $p'^*=p'-\frac{e_a-e_b}{2}ET$ and $\bar{p}^*=\bar{p}+\frac{e_a-e_b}{2}ET$ as abbreviations. Here the notation $\mathcal{AI}$ indicates the Airy-transformation

$$\mathcal{AI}(G(\omega);\omega)(\Omega) = \int\frac{d\omega}{\nu_E}\text{Ai}(\frac{\Omega-\omega}{\nu_E})G(\omega) \tag{A4}$$

where $\text{Ai}(x)$ is the conventional Airy function [28] and the scaling frequency $\nu_E$ is introduced in (29). As can be seen, the Airy transform will collapse to an identity transformation for zero fields, where $\nu_E$ tends to zero, and we obtain the standard $\mathcal{T}$-matrix equation provided we neglect the time retardation in the distribution functions. This result one would have been obtained without fields. In this way we derived a generalization of the Bethe-Salpeter equation including retardation effects in addition to the field dependence.

FIG. 1. The field-free triplet phase shifts for different densities vs. the energy difference of two correlated particles. For the low density limit the phase shift approaches the value $\pi$ indicating a bound state corresponding through the Levinson theorem. With increasing densities this bound state vanishes at a special critical density known as Mott transition. At higher densities the Pauli-blocking effects cause a sign change in the the phase shifts at small relative momenta equivalent to an effective repulsion. This leads to an onset of a resonance at energies around $2\mu$ which can be interpreted as the onset of superfluidity.

FIG. 2. The field-dependent triplet phase shifts for different densities vs. the energy difference of two correlated particles. The applied electric field strength are $20 \times E_c$. The same parameters are used as in Fig 1. The bound state at low densities turned into a resonance owing to the field effect and resulting in a value of the phase shift between 0 and $\pi$ at small relative energies.

FIG. 3. The field dependent triplet phase shifts (solid line) for 3 different densities together with the field free shifts (dotted line) vs the energy difference of two correlated particles. The applied electric field strength are $20 \times E_c$ corresponding to Fig 2. At small densities (top picture) the bound state is turned into a resonance resulting in a value of the phase shift between 0 and $\pi$. Further oscillations occur at higher energies due to field effects and the oscillating spectral function. At densities where the Pauli-blocking effects cause a repulsion in the field free case (middle picture) the field effects lead to a decreasing of Pauli-blocking. This is due to the enlargement of the available phase space by the external field. At high densities where the onset of the superfluid phase transition is starting as a resonance around energies of $2\mu$ (bottom picture) the field effects cause an abrupt onset of the coherence. Without fields this behaviour is found at smaller temperatures, which means that the pairing occurs already at higher temperatures due to the field.

FIG. 4. The field independent triplet phase shifts for 3 different temperatures chosen at two different density situations. The BCS resonance becomes enhanced with decreasing temperature. For nuclear matter density $n_o$ the critical temperature is about $5 MeV$. This illustrates the statement of the Fig 3 (bottom).

FIG. 5. The imaginary versus real parts of the bound state energy of the deuteron state vs. density for a temperature of 10 MeV for the field free case (above) and for a field of $20 E_c$ (below). Whereas without fields (above) the bound state vanishes at a special density and the imaginary part is zero, at high fields the imaginary part is finite indicating a finite lifetime of the correlated state and shows a complicated density dependence of the resonance energy and damping.

FIG. 6. The imaginary (below) and the real part (above) of the correlation-state energy for vanishing density vs. the applied electric field. The vanishing damping or infinite life time of the deuteron becomes finite for higher field strengths and the bound state energy turns into a resonance moving in the complex plane. At special field strengths it crosses even the real axes and turns into an oscillating behaviour which is not plotted here.

FIG. 7. The imaginary (below) and the real parts (above) of the correlation-state energy vs. the applied electric field for a density of $n_o/4$ corresponding to the situation $f$ of Fig 1. In contrast to Fig 6 a density is chosen where no deuterons are present without fields. One sees that the field dependence leads to a decreasing of the lifetime corresponding with an increasing of the imaginary part of the resonance energies at higher fields. It is pronounced that the small field behaviour shows a negative damping indicating an unstable situation. This is due to the fact that at densities of this high no deuterons are present anymore and the bulk nuclear matter becomes unstable against small perturbations in this density region.

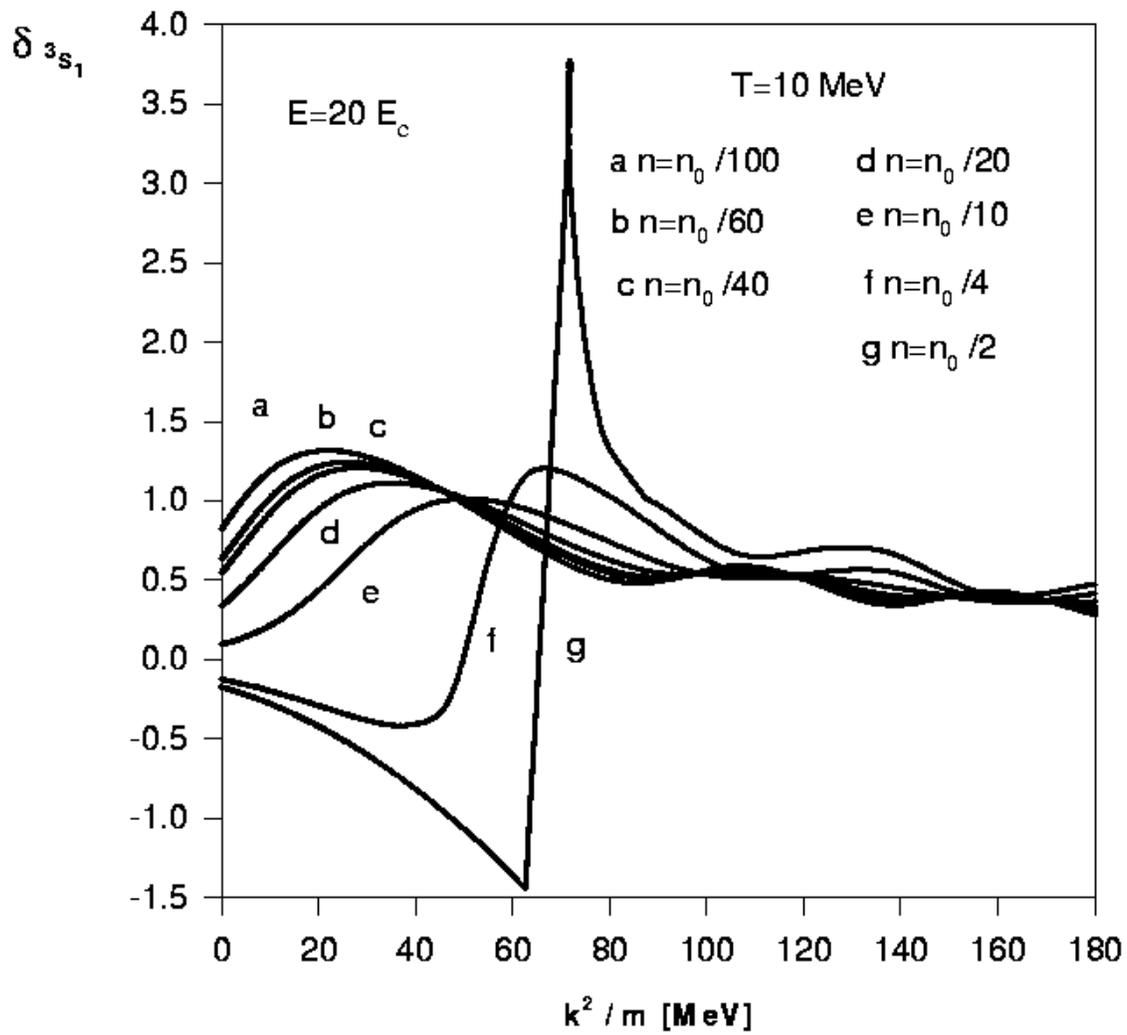

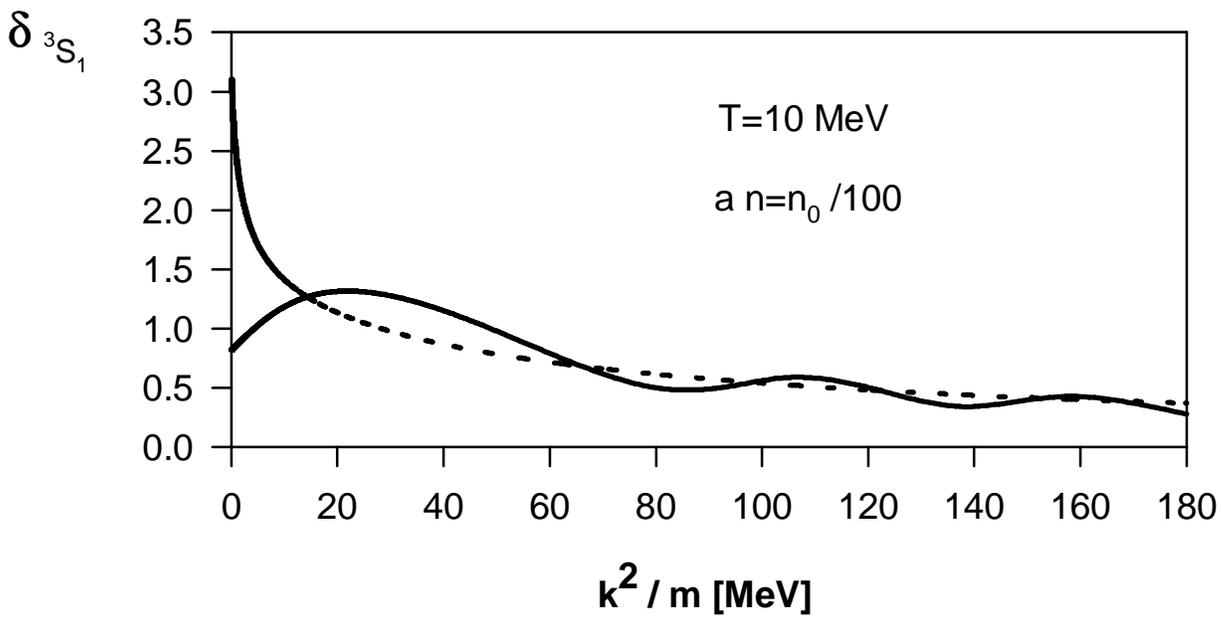

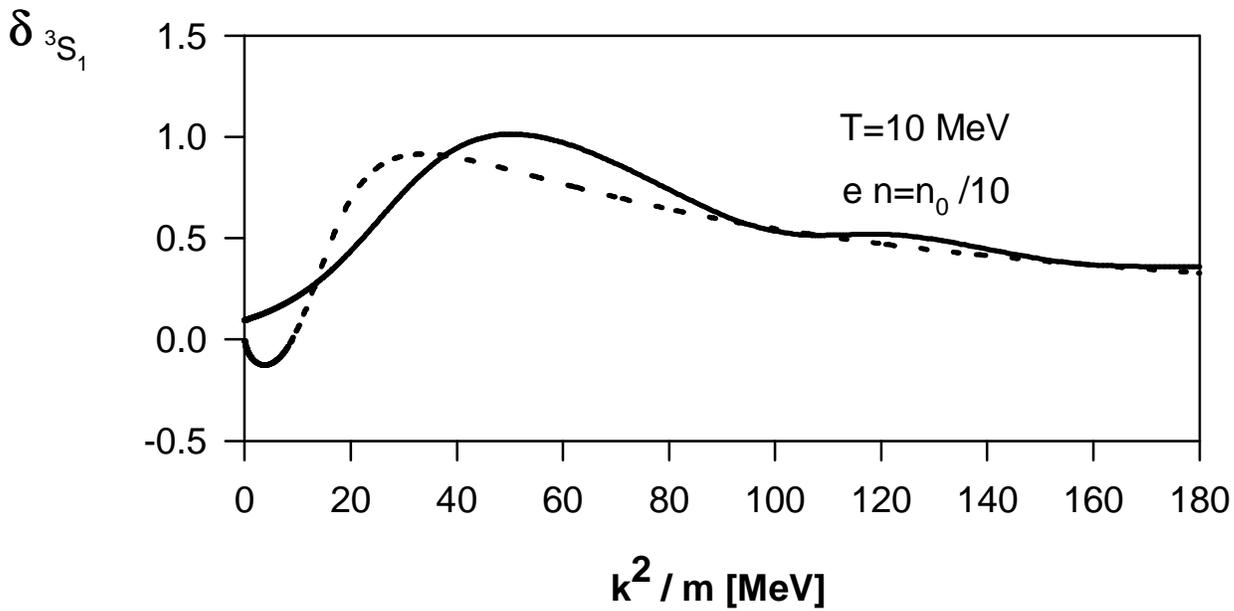

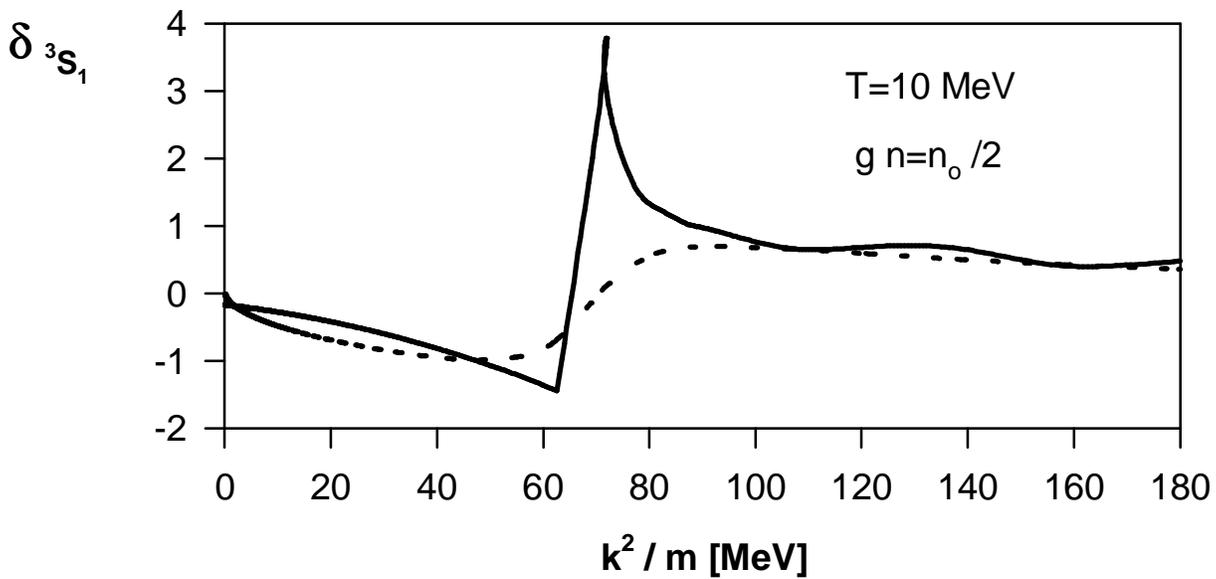

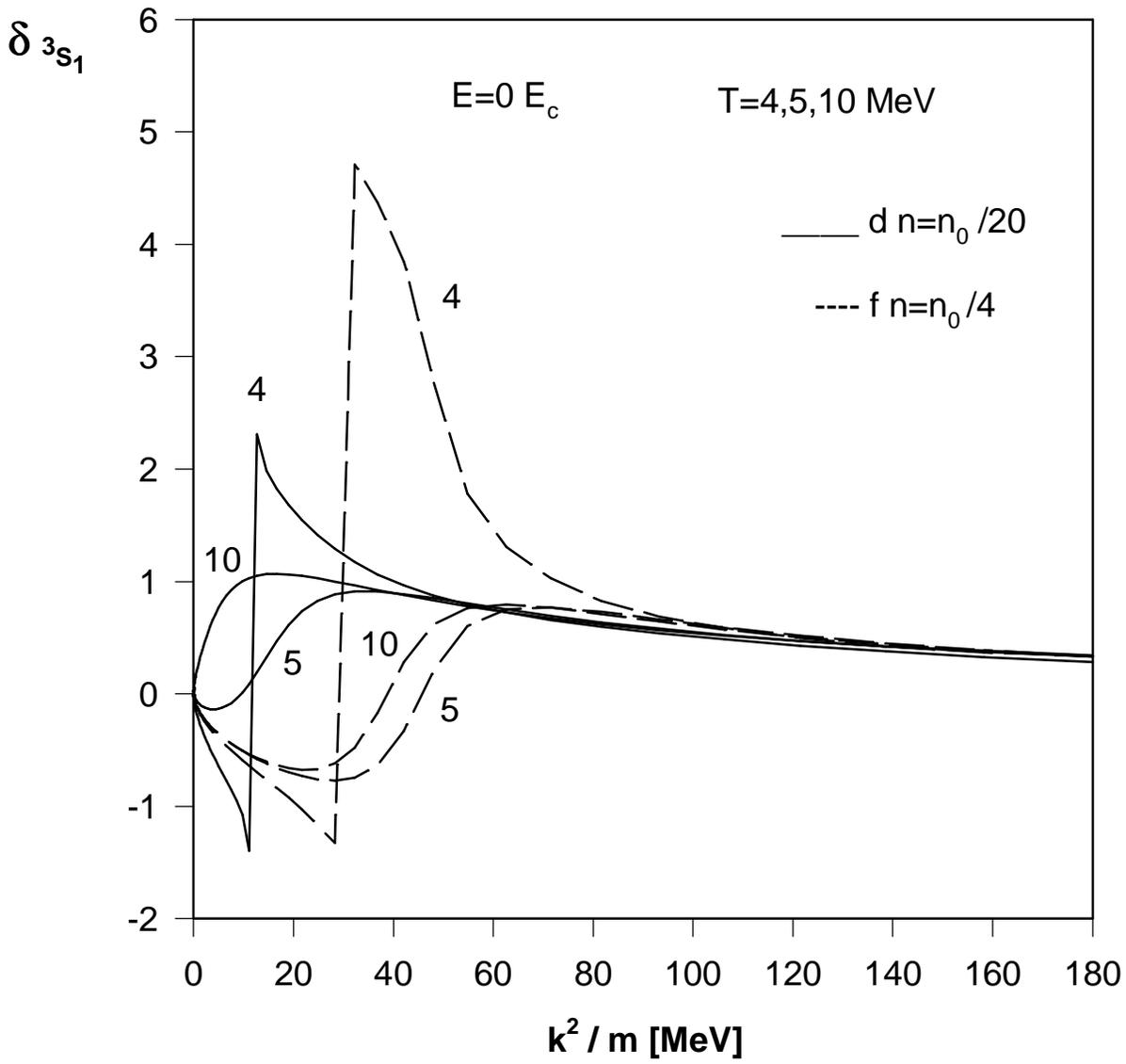

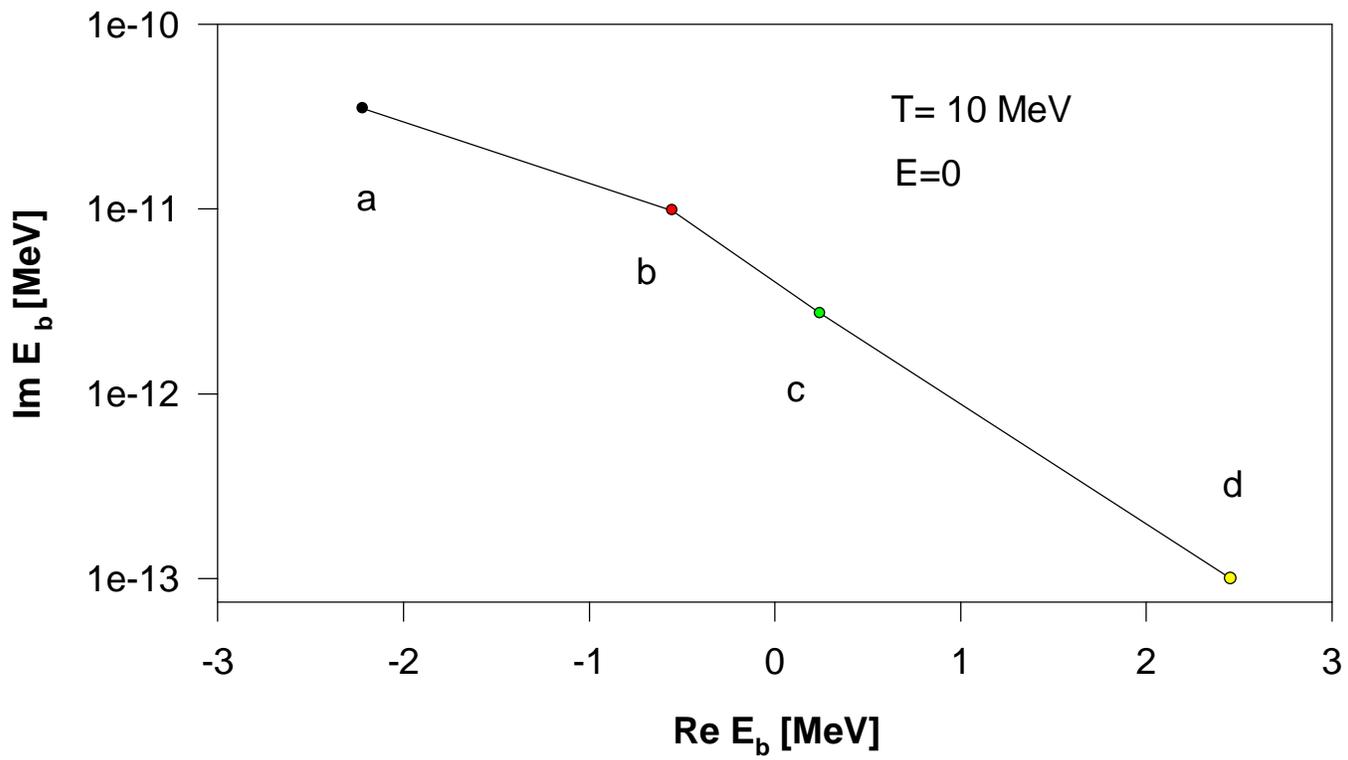

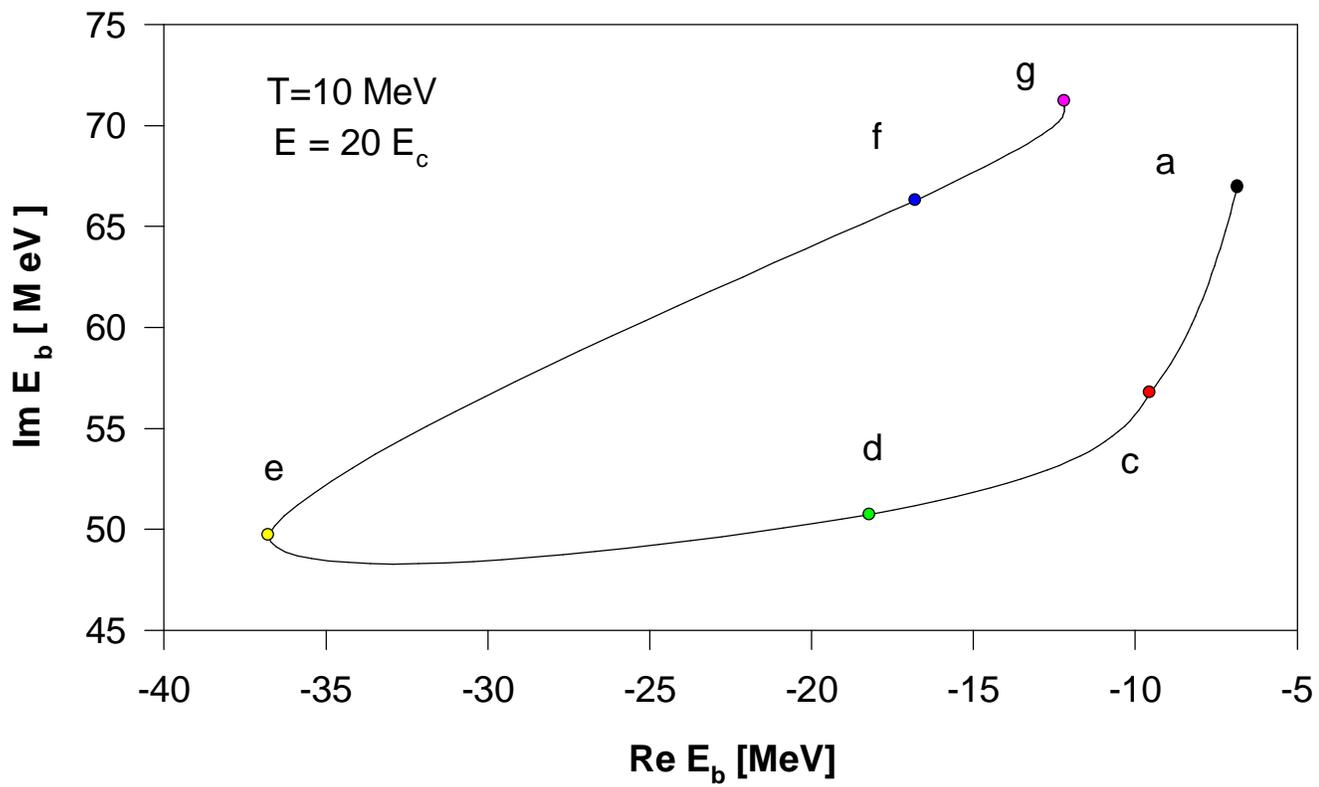

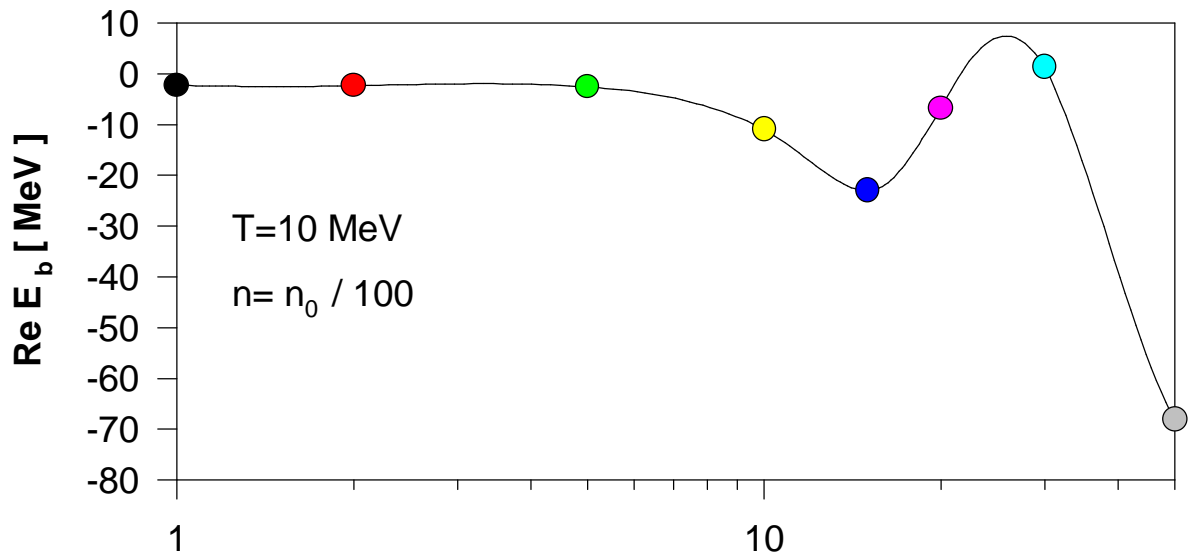

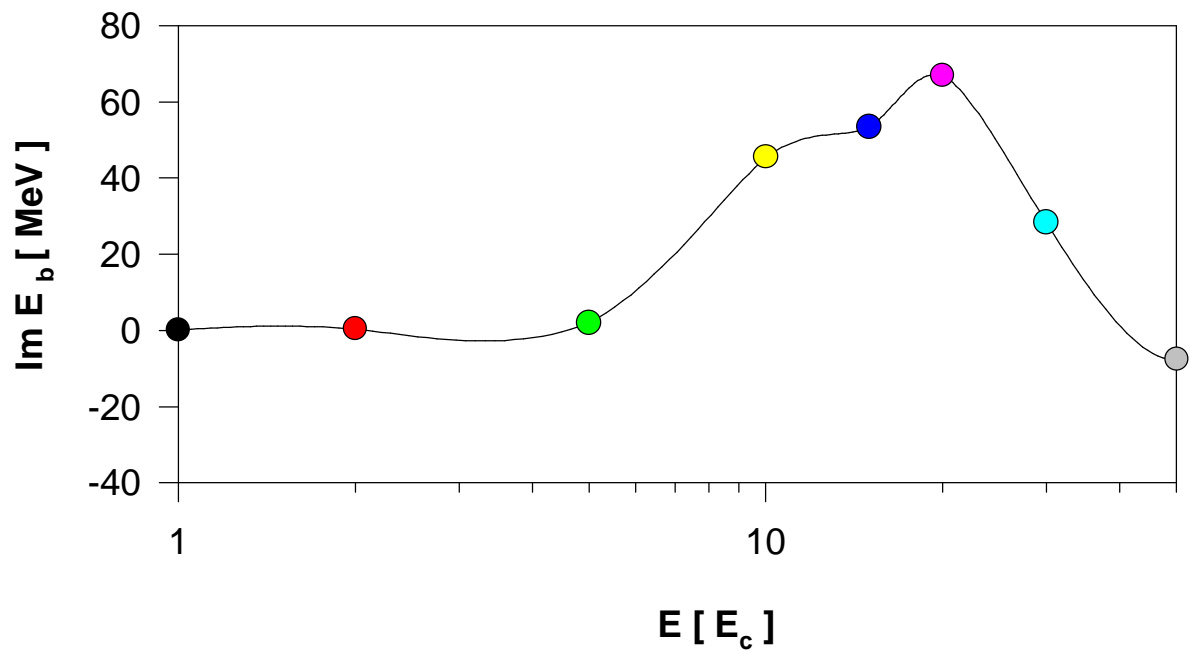

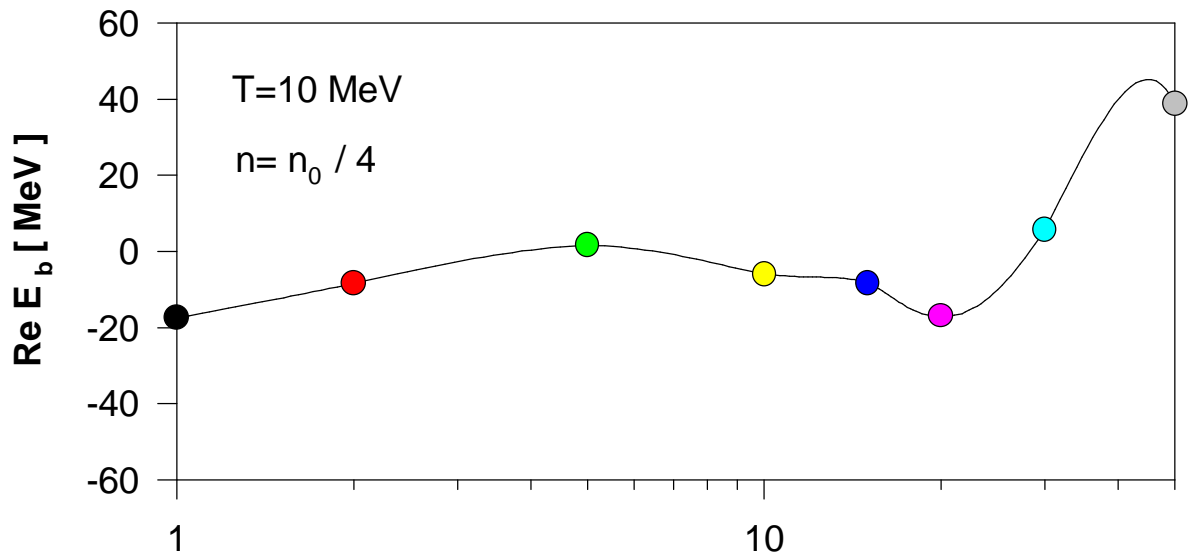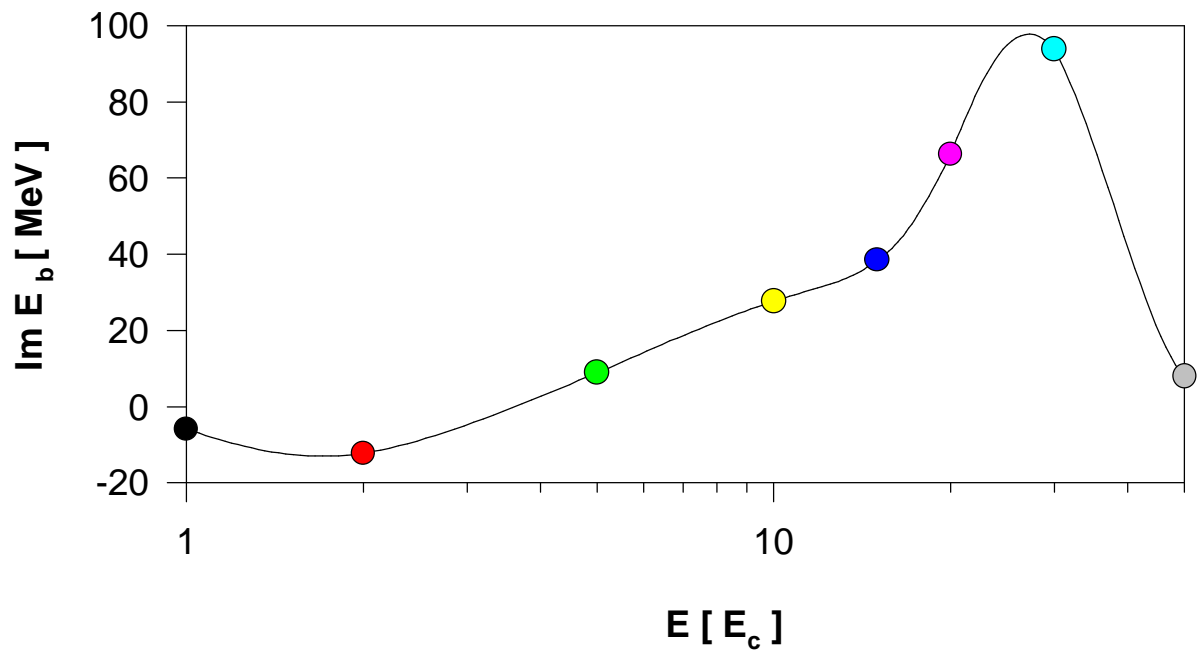